\documentclass{article}
\usepackage{spconf,amsmath,graphicx}
\usepackage{graphicx}
\usepackage{amssymb, amsmath, bm, mathtools,array}
\usepackage{textcomp}
\usepackage{hyperref}
\usepackage{verbatim}
\usepackage{lipsum} 
\usepackage{multirow}
\usepackage{siunitx}
\usepackage{diagbox}
\usepackage{dsfont}
\usepackage{colortbl}
\usepackage{booktabs}
\usepackage{caption}
\usepackage{subcaption}
\usepackage{algorithm}
\usepackage[noend]{algpseudocode}

\RequirePackage{color}\definecolor{BLUE}{rgb}{0,0,0}
\providecommand{\BLUE}[1]{{\color{BLUE}{#1}}}

\newcommand{\mIshort}{{${\text{\texttt{AL}}_{\texttt{NegE}}}$}}
\newcommand{\mIIshort}{{${\text{\texttt{AL}}_{\texttt{Adv}}}$}}
\newcommand{\mIIIshort}{{${\text{\texttt{AL}}_{\texttt{PosE}}}$}}
\newcommand{\mIVshort}{{${\text{\texttt{AL}}_{\texttt{Pas}}}$}}
\newcommand{\mVshort}{{${\text{\texttt{AL}}_{\texttt{Rem}}}$}}

\newcommand{\baseline}{{\texttt{Base}}}
\newcommand{\topline}{{\texttt{Top}}}
\newcommand{\setA}{{\mathtt{A}}}
\newcommand{\setD}{{\mathtt{B}}}

\title{Investigating Active-learning-based Training Data Selection \\ for Speech Spoofing Countermeasure}
\name{Xin Wang\thanks{This study is supported by JST CREST Grants (JPMJCR18A6 and JPMJCR20D3), MEXT KAKENHI Grants (21K17775, 21H04906), and Google AI for Japan program.}, Junichi Yamagishi}
\address{National Institute of Informatics, Japan }

\begin{document}
%\ninept
%
\maketitle

\fbox{%
\footnotesize
  \parbox{0.4\textwidth}{
Copyright 2023 IEEE. Published in the 2022 IEEE Spoken Language Technology Workshop (SLT) (SLT 2022), scheduled for 19-22 January 2023 in Doha, Qatar. Personal use of this material is permitted. However, permission to reprint/republish this material for advertising or promotional purposes or for creating new collective works for resale or redistribution to servers or lists, or to reuse any copyrighted component of this work in other works, must be obtained from the IEEE. Contact: Manager, Copyrights and Permissions / IEEE Service Center / 445 Hoes Lane / P.O. Box 1331 / Piscataway, NJ 08855-1331, USA. Telephone: + Intl. 908-562-3966.
  }%
}

\begin{abstract}
Training a speech spoofing countermeasure (CM) that generalizes to various unseen test data is challenging. Methods such as data augmentation and self-supervised learning can help, but the imperfect CM performance still calls for additional strategies. This paper investigates CM training using active learning (AL) to select useful training data from a large pool set, which is an unexplored area for speech anti-spoofing. Existing AL methods are compared to select useful data from a large pool set. A new AL method is also proposed that actively removes useless data from a pool. Experiments demonstrate that an energy-score-based AL method and the proposed data-removing method outperformed our strong baseline, and the relative reduction in detection error rates was higher than 40\% on multiple test sets. Furthermore, compared with a top-line method that blindly used the whole pool set for training, the two AL-based CMs used less training data and achieved better or similar performance.
\end{abstract}

\section{Introduction}
Detection of spoofed speech signals synthesized through text-to-speech (TTS) and voice conversion (VC) technologies is now a well-established topic \cite{evans2013spoofing,wu2017asvspoof}. It is usually formulated as a machine learning task that consists of feature extraction and classifier training to discriminate the spoofed and bona fide data (i.e., real human voices) in a training set. It further requires the trained model, or the so-called spoofing countermeasure (CM), to reliably classify incoming trials even if the trials are from unknown TTS/VC systems or uttered by unseen speakers.  In other words, the CM should be generalizable to unseen data.

This challenging task has given rise to many tentative solutions involving feature extraction algorithms \cite{davis1980comparison, Todisco2017, kamble_sailor_patil_li_2020}  and various shallow or deep-neural-network (DNN)-based classifiers \cite{he2016deep, Lavrentyeva2017}.
Meanwhile, databases for speech anti-spoofing have been constructed \cite{wu2015asvspoof, korshunov2016overview, Todisco2019, frank2021wavefake, zhang2021fmfcc}. To simulate real applications, these databases have disjoint training and test sets, and the test sets contain unseen attacks and speakers.   
Although state-of-the-art CMs have performed impressively well on standard databases, 
it was also reported that CMs well trained on one database made significantly more errors on the test set of a different database \cite{paul2017generalization,das2020assessing,muller21_asvspoof,wang2021investigating,muller2022does}.

Such degradation may be caused by the different languages, channel variations, or even artifacts in the training set \cite{muller21_asvspoof}. This is unsurprising since the training set of a standard anti-spoofing database is relatively small and derived from a single source speech database. 
Therefore, it is reasonable to consider approaches that expand the training data space. One good example is data augmentation, a method that augments training sets using waveforms processed with codec or other signal processing operators \cite{Chen2020Odyssey, das2021data, Tak2021}. 
Another approach is to replace conventional feature extractors with a pre-trained self-supervised-learning (SSL) DNN \cite{wang2021investigating,tak2022automatic}. This DNN is pre-trained on a huge amount  of diverse bona fide speech data \cite{NEURIPS2020_92d1e1eb}, and it is able to extract features robust to various channel conditions, languages, and speakers. 

Another yet unexplored direction is to add more data to the CM training set, hoping that the newly added data can cover unseen factors.  Although this approach seems to be obvious, it is inefficient to simply add all the data available to the training set.  For one thing, more training data requires more training time. Furthermore, not all the data are useful to a CM, and some may be detrimental. 
Therefore, how to select useful training data and avoid those harmful is an intriguing research question.

\begin{figure}[t]
\centering
\includegraphics[trim=0 360 0 80, clip, width=0.9\columnwidth]{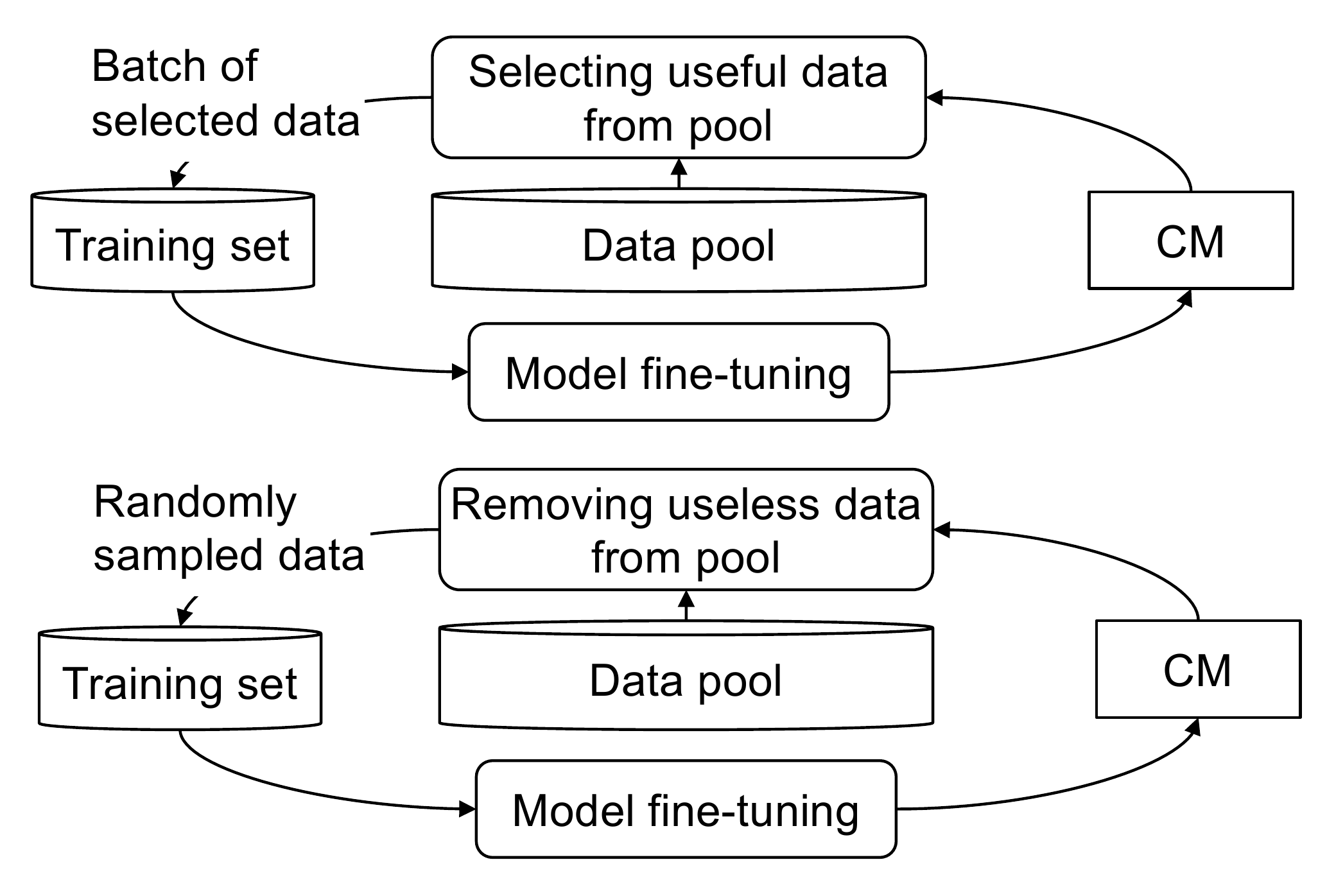}
\vspace{-2mm}
\caption{Illustration of AL-based training frameworks for spoofing CM. Top figure is standard AL framework. Bottom figure is proposed framework that removes useless data.} 
\vspace{-5mm}
\label{fig:idea}
\end{figure}

To address the question, this study investigated active learning (AL) for CMs, a framework that selects useful data and fine-tunes CMs in an iterative manner \cite{settles.tr09}. The idea is illustrated in Figure~\ref{fig:idea}.  
Given candidate data in a so-called pool set, AL asks the CM to evaluate the usefulness of each piece of candidate data and selects the most useful ones to expand the training set. It then fine-tunes the CM before the next iteration.  
This study compares a few existing algorithms to measure the usefulness of candidate data, including one based on an energy score and another using adversarial samples. 
\BLUE{Furthermore, a variant of the AL framework is proposed that removes useless data from a pool set before randomly sampling new data to expand the training set.}

In this study, thorough experiments are conducted to measure the effectiveness of the above AL frameworks on multiple test sets including ASVspoof 2019 \cite{Todisco2019}, 2021 \cite{yamagishi21_asvspoof}, and another database called WaveFake \cite{frank2021wavefake}. Given a large and diverse pool set, a negative-energy-score-based AL method and the proposed data-removing method are shown to outperform a strong baseline. Furthermore, the two methods are more efficient than a naive top-line method that simply merges the pool set with the training set. The AL methods use less than one-fourth of the candidate data and achieve roughly the same equal error rate (EER) as the top-line method.

This study has two main contributions. First, AL is known to be difficult to tune and use \cite{attenberg2011inactive, cawley2011baseline, chandra2021initial}, and no recipe is available for spoofing CMs. This study compared existing AL methods and summarizes useful practices for AL-based CMs.  Second, this study describes a new AL variant that actively removes useless data. The findings are encouraging and expected to be useful to the research field.

\section{AL-based training framework for CM}
\label{sec:method}

\subsection{Active Selection of Useful Data}
AL is a machine-learning research topic with a relatively long history \cite{MacKay1992,cohn1996active}. The general goal is to allow a model \emph{to select its own training data}\cite{cohn1996active}. 
For a spoofing CM, although a large amount of bona fide and spoofed data can be collected from the Internet or created using open-sourced tools in advance, simply pooling all the data may be problematic because the pool may be imbalanced in terms of speakers, spoofing methods, or languages. It is more efficient to select only a useful subset from a large pool to train a CM. Such a goal can be achieved using the pool-based AL framework \cite{settles.tr09}.

\subsubsection{General framework}
\textbf{Algorithm 1} describes the pool-based AL framework used in this study. It starts with a relatively small seed set $\mathcal{U}_{\text{seed}}$ and trains the CM from scratch. It then enters the AL training loop, where each iteration selects $L$ data from a large pool set $\mathcal{U}_{\text{pool}}$. The selected data are removed from the pool and added to the training set. The CM is then fine-tuned on the expanded training set before the next iteration. 

For a spoofing CM, we assume that $\mathcal{U}_{\text{seed}}$ is from a well-designed database for speech anti-spoofing. In contrast, $\mathcal{U}_{\text{pool}}$ is a collection of speech data from various sources.
\textit{It may contain valuable data not covered by $\mathcal{U}_{\text{seed}}$, but it may also contain biased, noisy, or redundant data.} 
The AL framework is expected to actively select useful data from $\mathcal{U}_{\text{pool}}$.

The key question then is how to measure data usefulness (i.e., line 5 in \textbf{Algorithm~\ref{alg:cap}}). 
While many strategies exist, a widely used category is the so-called certainty scoring \cite{settles.tr09}. 
The assumption is that a datum is more useful if the CM is less certain to classify it. 
Therefore, by adding data with a smaller certainty score to the training set, the CM is expected to learn a better decision boundary to classify similarly difficult data. 

\begin{algorithm}[t]
\footnotesize
\caption{Active data selection for AL-based CM training}\label{alg:cap}
\begin{algorithmic}[1]
\Require Seed set $\mathcal{U}_{\text{seed}}\leftarrow \{\boldsymbol{X}_1, \cdots, \boldsymbol{X}_N\}$
\Require Pool set $\mathcal{U}_{\text{pool}}\leftarrow \{\boldsymbol{X}_{N+1}, \cdots, \boldsymbol{X}_{N+M}\}, N \ll M$

\noindent Note that $\boldsymbol{X}_n=\{\boldsymbol{o}_{1:T_n}, y_n\}$, where $\boldsymbol{o}_{1:T_n}$ is a waveform of length  $T_n$, and $y_n \in\{\text{Bona fide}, \text{Spoof}\}$ is a label. 
\State $\mathcal{U}_{\text{train}} \leftarrow \mathcal{U}_{\text{seed}}$
\State CM $\leftarrow \text{Training-from-scratch}(\mathcal{U}_{\text{train}})$
\Repeat 
\For{$\boldsymbol{X}_m \in \mathcal{U}_{\text{pool}}$}
\State $c_m = \mathcal{F}_{\text{CM}}(\boldsymbol{o}_{1:T_m}, y_m, \mathcal{U}_{\text{train}})$ \Comment{Certainty scoring}
\EndFor
\State $\mathcal{I} \leftarrow \text{argmin-sort}_{m}(\{\cdots, c_{m}, \cdots\})[0:L]$ 
\State $\mathcal{V}_{\text{useful}} \leftarrow \{\boldsymbol{X}_m \in \mathcal{U}_{\text{pool}} | m \in \mathcal{I}\}$ \Comment{Retrieve $L$ data with smallest certainty scores}
\State $\mathcal{U}_{\text{pool}} \leftarrow \mathcal{U}_{\text{pool}} \setminus  \mathcal{V}_{\text{useful}}$ \Comment{Remove from pool }
\State $\mathcal{U}_{\text{train}} \leftarrow \mathcal{U}_{\text{train}} \cup \mathcal{V}_{\text{useful}}$ \Comment{Expand training set}
\State CM $\leftarrow \text{Fine-tuning}(\text{CM}, \mathcal{U}_{\text{train}})$ \Comment{Fine-tune CM}
\Until{$K$ iterations are completed}
\end{algorithmic}
\end{algorithm}

This study compares a few of the certainty scoring methods, and the details are given in the next subsection. 
All these methods compute a certainty score $c_m\in\mathbb{R}$ for each utterance in the pool set. Note that the notation in line 5 of Algorithm~\ref{alg:cap} is for a general definition. The scoring function $\mathcal{F}_{\text{CM}}$ may not necessarily use $y_m$ or  $\mathcal{U}_{\text{train}}$ in implementation.

\subsubsection{Certainty scoring methods}
\label{sec:method-cm}

\noindent \textbf{Negative energy-based certainty score}:
The first method is inspired by a related work on a DNN-based CM \cite{Wang2021}. 
Given an input waveform $\boldsymbol{o}_{1:T_m}$, the CM extracts features and transforms them through multiple hidden layers and a softmax output layer for binary classification. Let the input logits to the softmax be $\{l_{m,1}, l_{m,2}\}$. Then, the certainty score $c_m$ is computed by
$c_m = \mathcal{F}_{\text{CM}} (\boldsymbol{o}_{1:T_m}) = -\log\sum_{j=1}^{2}\exp(l_{m,j})$.

The computed $c_m$ is also referred to as a negative energy score \cite{NEURIPS2020_f5496252}, and it has been demonstrated that $c_m$ tends to be smaller when the CM is less certain of the input data \cite{Wang2021}. 
Furthermore, it has outperformed the classical entropy-based certainty score\footnote{\BLUE{For certainty scoring, using the entropy of output probabilities for a binary classifier is equivalent to using the max-output-probability in \cite{Wang2021}.}} for CMs \cite{Wang2021}.
Therefore, it is selected as a candidate scoring method in this study.

\vspace{2mm}
\noindent \textbf{Adversarial-sample-based distance}:
\label{sec:adv}
This method assumes that useful data can be found near the adversarial samples to the CM.
It is included in this study for comparison because it is a relatively new scoring method and has performed well for AL \cite{mayer2020adversarial} on image classification. 

Given a mini-batch of training data randomly drawn from $\mathcal{U}_{\text{train}}$, this method produces adversarial samples by back-propagating gradients through the CM to the input waveforms \cite{mayer2020adversarial}. It then computes the distances between $\boldsymbol{o}_{1:T_m}$ from the pool set and the generated adversarial samples. The distance to the closest adversarial sample is used as the certainty score for $\boldsymbol{o}_{1:T_m}$. The algorithm is described below. \vspace{-2mm}
\begin{algorithm}
\footnotesize
\begin{algorithmic}[1]
\Function{$c_m=\mathcal{F}_{\text{CM}}$}{$\boldsymbol{o}_{1:T_m}, \mathcal{U}_{\text{train}}$}
\ForAll {$b\in\{1, \cdots, H\}$} 
\State $ \{\boldsymbol{o}_{1:T_h}, y_h\} \leftarrow\text{Random-select}(\mathcal{U}_{\text{train}})$
\State $\tilde{\boldsymbol{o}}_{1:T_h} \leftarrow\text{Adv.Gen}(\boldsymbol{o}_{1:T_h}, y_h, \text{CM})$ 
\EndFor
\State $c_m = \min_{h} || \text{NN}_{\text{CM}}(\boldsymbol{o}_{1:T_m}) - \text{NN}_{\text{CM}}(\tilde{\boldsymbol{o}}_{1:T_h}) ||_2$ 
\EndFunction
\end{algorithmic}
\end{algorithm}
\vspace{-2mm}

\noindent Note that $H$ is equal to the mini-batch size for CM training. $\text{NN}_{\text{CM}}(\boldsymbol{o})$ is the output vector from a temporal pooling layer in our CMs. It has a fixed number of dimensions.  

\vspace{2mm}
\noindent \textbf{Random scoring:}
The third method produces $c_m$ by simply drawing a random number from a uniform distribution between 0 and 1. This method is equivalent to  random sampling from a pool set. Since it does not use the knowledge from a CM during the data selection process, it is also referred to as \emph{passive learning} \cite{settles.tr09}. Despite its simplicity, it has been found to be effective in many research fields  \cite{chandra2021initial, Holub2008, ducoffe2018adversarial,shui2020deep}.

\subsection{{Active Removing of Useless Data}}
With a proper scoring method, the aforementioned AL framework selects the ``most useful'' data per iteration. Such a greedy exploitation strategy may introduce sampling bias \cite{dasgupta2009two} and lead to a local minimum. Exploration, i.e., selecting data that are not the most useful, may bring in more data diversity. 

However, exploration over an entire pool set is unnecessary because some data cannot add new information to the CM and are useless.  
To make exploration more efficient, we propose an AL framework that actively removes useless data before exploration in each iteration. \textbf{Algorithm 2} lists the procedures. It is similar to \textbf{Algorithm 1} but has a few modifications (highlighted in red color). In this study, we use the negative energy score to measure the data usefulness and the random sampling method for exploration
\footnote{It is possible to select the most useful data in line 9 of \textbf{Algorithm 2}. This method is slightly different from Algorithm 1 because it will remove both ${\mathcal{V}_{\text{batch}}}$ and ${\mathcal{V}_{\text{useless}}}$ in every iteration. However, this exploitation-only method did not improve the performance.}.

\begin{algorithm}[t]
\footnotesize
\caption{Active data removing for AL-based CM training}\label{alg:cap2}
\begin{algorithmic}[1]
\Require Seed set $\mathcal{U}_{\text{seed}}\leftarrow \{\boldsymbol{X}_1, \cdots, \boldsymbol{X}_N\}$
\Require Pool set $\mathcal{U}_{\text{pool}}\leftarrow \{\boldsymbol{X}_{N+1}, \cdots, \boldsymbol{X}_{N+M}\}, N \ll M$
\State $\mathcal{U}_{\text{train}} \leftarrow \mathcal{U}_{\text{seed}}$
\State CM $\leftarrow \text{Training-from-scratch}(\mathcal{U}_{\text{train}})$
\Repeat 
\For{$\boldsymbol{X}_m \in \mathcal{U}_{\text{pool}}$}
\State $c_m = \mathcal{F}_{\text{CM}}(\boldsymbol{o}_{1:T_m}, y_m, \mathcal{U}_{\text{train}})$ \Comment{Certainty scoring}
\EndFor
\State $\mathcal{I} \leftarrow \text{\textcolor{red}{argmax-sort}}_{m}(\{\cdots, c_{m}, \cdots\})[0:L]$ 
\State $\mathcal{V}_{\text{useless}} \leftarrow \{\boldsymbol{X}_m \in \mathcal{U}_{\text{pool}} | m \in \mathcal{I}\}$ 
\State $\textcolor{red}{\mathcal{U}_{\text{pool}} \leftarrow \mathcal{U}_{\text{pool}} \setminus  \mathcal{V}_{\text{useless}}}$ \Comment{\textcolor{red}{Remove useless data}}
\State $\textcolor{red}{\mathcal{V}_{\text{batch}} \leftarrow \text{Random-select}(\mathcal{U}_{\text{pool}})}$  \Comment{Random selection}
\State $\mathcal{U}_{\text{pool}} \leftarrow \mathcal{U}_{\text{pool}} \setminus  \mathcal{V}_{\text{batch}}$ 
\State $\mathcal{U}_{\text{train}} \leftarrow \mathcal{U}_{\text{train}} \cup \mathcal{V}_{\text{batch}}$ 
\State CM $\leftarrow \text{Fine-tuning}(\text{CM}, \mathcal{U}_{\text{train}})$
\Until{$K$ iterations are completed}
\end{algorithmic}
\end{algorithm}

\section{Experiments}
\label{sec:ex}

\subsection{Data and Protocols}
We first collected diverse speech data to compose the AL seed and pool sets.  
As shown in Table~\ref{tab:source-data} (a),  subsets \textcircled{1} and \textcircled{2} are the training sets of the ASVspoof 2019 LA \cite{Todisco2019} and FMFCC-A \cite{zhang2021fmfcc} anti-spoofing databases, respectively.  \textcircled{3} and \textcircled{4} were constructed using ESPNet TTS models \cite{hayashi2020espnet} pre-trained on the LibriTTS \cite{zen2019libritts} and LJSpeech \cite{ljspeech17} TTS databases, respectively. \textcircled{5} was sourced from Blizzard Challenge (BC) 2019 \cite{wu2019blizzard}. These data subsets have varied languages, speakers, and recording conditions, and they cover many recent TTS models and neural vocoders. 
Last, a small subset from VoxCeleb 1 \cite{Nagrani2020} was also included.

With the above data subsets, we composed the AL seed and pool sets listed in Table~\ref{tab:source-data} (b). 
The seed set was the ASVspoof 2019 LA training set, and it simulated the common scenario where the seed set is from a standard anti-spoofing database. 
Two pool sets were constructed with varied degrees of diversity.  Pool set $\setA$ simulated the case where spoofing trials can be collected from multi-speaker TTS systems, 
while set $\setD$ is a more diverse case that covers more speakers, attacks, and languages.

During evaluation, we used multiple test sets to compute EERs. Following \cite{yamagishi21_asvspoof}, we used the test sets from the 2019 LA, 2021 LA, and 2021 Deepfake (DA) evaluation tracks. We also included the test set of WaveFake \cite{frank2021wavefake} as more challenging test data.

\begin{table}[!t]
\caption{Base data subsets, AL seed set, and pool sets used in this study. 
Five columns on right side list number of trials (bona fide/spoof), duration, languages, number of speakers, and number of spoofing attacks (i.e., types of TTS and VC algorithms), respectively.} 
\vspace{-5mm}
\begin{center}
\setlength{\tabcolsep}{3pt}
\resizebox{\columnwidth}{!}{
\begin{tabular}{rrrrrrrr}
\multicolumn{7}{c}{(a) {Base data subsets}}\\
\toprule
                ID        & Source          & \#. Trial & Dur. (h) & Lang. & \#. Spr & \#. Att\\
\midrule 
  \textcircled{1} & ASVspoof2019 LA trn. & 2,580 / 22,800 & 24.0 & En & 20 & 6\\ 
 \textcircled{2} & FMFCC-A, trn. & 4,000 / \phantom{0}6,000 & 5.5 & Zh & 77 & 5 \\
   \textcircled{3} & ESPNet on LibriTTS&  \phantom{0,}736 / \phantom{0}4,275 & 8.0 & En & $>$80 & 6 \\
 \textcircled{4} & ESPNet on LJSpeech & \phantom{0,}200 / \phantom{0}1,800  & 3.6 & En & 1 & 9 \\
  \textcircled{5} & BC 2019 &  \phantom{00,}75 / \phantom{0}5,925 & 15.5 & Zh & 1 & 25 \\
  \textcircled{6} & VoxCeleb1 &  6,000 / \phantom{00,00}0 & 13.6 & Mul. & $>$1 k & 0\\
\bottomrule
\multicolumn{6}{c}{\phantom{0}}\\

\multicolumn{6}{c}{(b) {AL seed and pool sets created from base data subsets}}\\
\toprule
 & {Base data} & \#. Trial & Dur. (h) & Lang. & \#. Spr & \#. Att. \\
\midrule 
Seed set & \textcircled{1} & 2,580 / 22,800 & 24.0 & En & 20 & 6\\ 
Pool set $\setA$ & \textcircled{2}+\textcircled{3} & \phantom{0}4,736 / 10,275 & 13.5 & En, Zh & $>$150 & 11 \\
Pool set $\setD$ & {\textcircled{2}+\textcircled{3}+\textcircled{4}+\textcircled{5}+\textcircled{6}} & 11,011 / 18,000 & 46.3 & Mul. & $>$1.1 k  & 45\\
\bottomrule 
\end{tabular}
}
\vspace{-6mm}
\end{center}
\label{tab:source-data}
\end{table}%

\subsection{Experimental Models and Configurations}
\label{sec:recipe}
The experiment included the following CMs:
\begin{itemize}
\setlength{\itemsep}{1pt}
  \setlength{\parskip}{0pt}
  \setlength{\parsep}{0pt}
\item \mIshort: active data selection based on the negative energy score;
\item \mIIshort: active data selection based on the distance to adversarial examples;
\item \mVshort: active data-removing method based on the negative energy score, followed by random sampling;
\item \mIVshort: passive data selection by random sampling;
\item \mIIIshort: same as \mIshort\ except that the certainty score was multiplied by -1.
\end{itemize}
Note that \mIIIshort\ was included as a reference system. It is theoretically the least effective method because it selects the useless trials from a pool set in each AL iteration. 
Additionally, two CMs using a non-AL training scheme were included:
\begin{itemize}
\setlength{\itemsep}{1pt}
  \setlength{\parskip}{0pt}
  \setlength{\parsep}{0pt}
\item \baseline\  was trained on the seed set only;
\item \topline\ was trained on the merged seed and pool sets.
\end{itemize}
\baseline\ represents common CMs trained on a standard anti-spoofing database, while 
\topline\ simulates the case where new data were simply merged with the seed set. 

All the CMs used the same backbone DNN architecture inspired by \cite{wang2021investigating}.
It consists of a SSL-based DNN called wav2vec 2.0 \cite{NEURIPS2020_92d1e1eb} as the feature extractor. 
The extracted feature sequence is merged into a single vector through temporal pooling (i.e., $\text{NN}_{\text{CM}}(\boldsymbol{o})$ in Section~\ref{sec:adv}). After that, it is propagated through linear transformation and the softmax to produce the binary output probability. During training, the SSL-based feature extractor was also updated.

All the CMs except \baseline\ were trained and compared using seed and pool set $\setA$ and $\setD$. 
\topline\ was also trained using the seed and pool sets, while \baseline\  was trained only on the seed set (i.e., ASVspoof 2019 LA training set). 
Note that this \baseline\ was the best performing CM in \cite{wang2021investigating}.

To accelerate the experiments, the trained \baseline\ was used as the seed model for the AL-based CMs (line 2 in Algorithms 1 and 2). The number of data selected $L$ per iteration was 2,560\footnote{
Due to limited GPU memory, all the waveforms in the pool were split into segments no longer than 4 s. What AL selected were 2,560 segments.}, and the maximum number of AL iterations $K$ was set to 8.  The AL-based CMs were fine-tuned for 5 epochs per iteration.  The choice of $L$ is application-dependent, and our pilot study found that an $L$ of less than 1,000  was ineffective. Given $L=2,560$, we set $K=8$ so that all the data in pool set $\setA$ were used up, i.e., $\mathcal{U}_{\text{pool}} =\varnothing$ at the end. This helped with the experiment analysis. Meanwhile, $K=8$ did not consume all the data in pool set $\setD$, which simulated real applications where there was a budget for the total number of data selected.

For fair comparison with the AL-based CMs, \topline\ was also initialized with \baseline\ and fine-tuned using the merged data set with early stopping \footnote{Early stopping was done using a development set. For each pool set, we prepared a development set that was randomly drawn from the same data source as that for the pool set, but these two sets were disjoint. The development set was used only for early stopping of \topline, which helped in avoiding under-fitting or over-fitting.}.
All the training and fine-tuning used the Adam optimizer with $\beta_1=0.9, \beta_2=0.999, \epsilon=10^{-8}$ \cite{kingma2014adam}, a mini-batch size of 16, and a learning rate of $1\times10^{-6}$. 
For fair evaluation,  we repeated the training and evaluation of each CM three times, each time using a different random initial state. 

\begin{figure*}[t]
       \centering
       \begin{subfigure}[t]{0.95\textwidth}
        \includegraphics[trim=0 25 0 0, clip, width=0.95\textwidth]{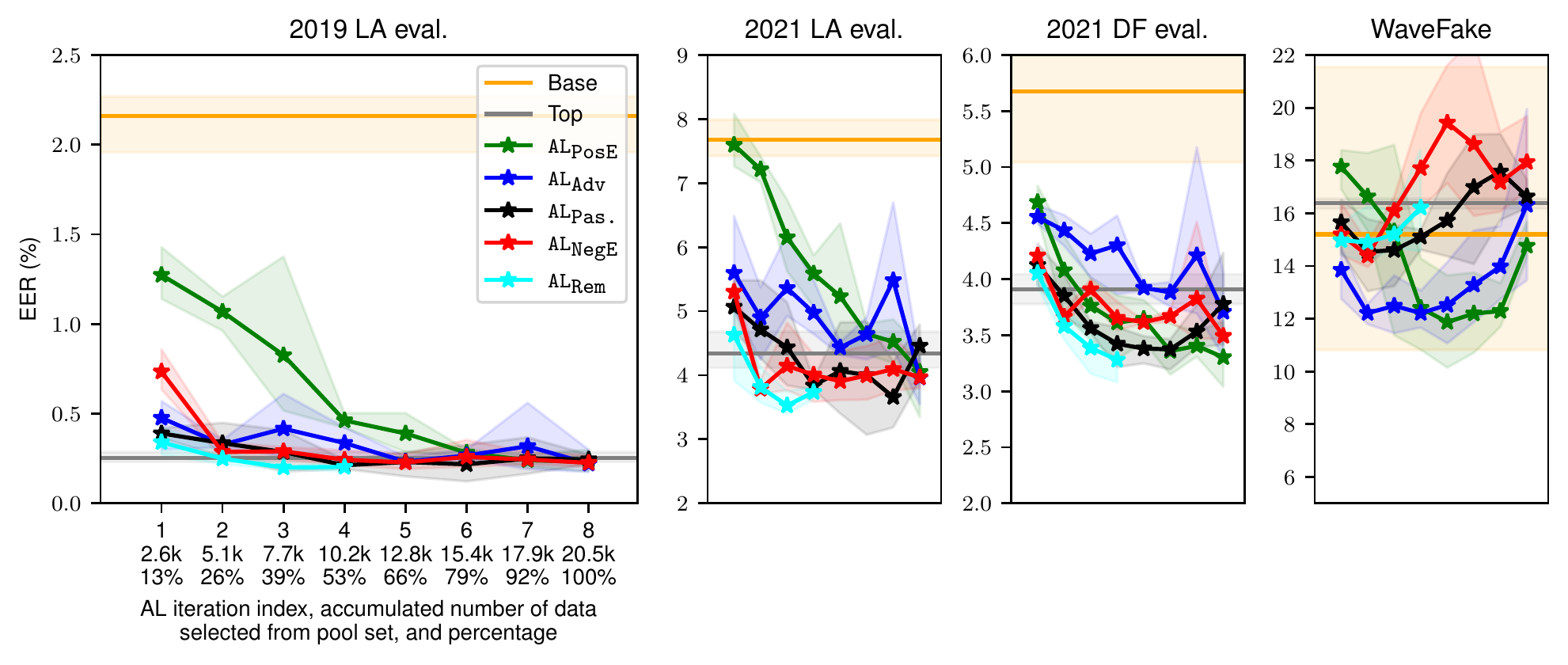}
        \label{fig:results_set1}
        \caption{Results with pool set $\setA$. Note that, after eight AL iterations, all data in pool set $\setA$ were added to training set. \mVshort\ used up all data after 4th iteration since it also removes useless data from pool. }
     \end{subfigure}
      \begin{subfigure}[t]{0.95\textwidth}
        \includegraphics[trim=0 25 0 0, clip, width=0.95\textwidth]{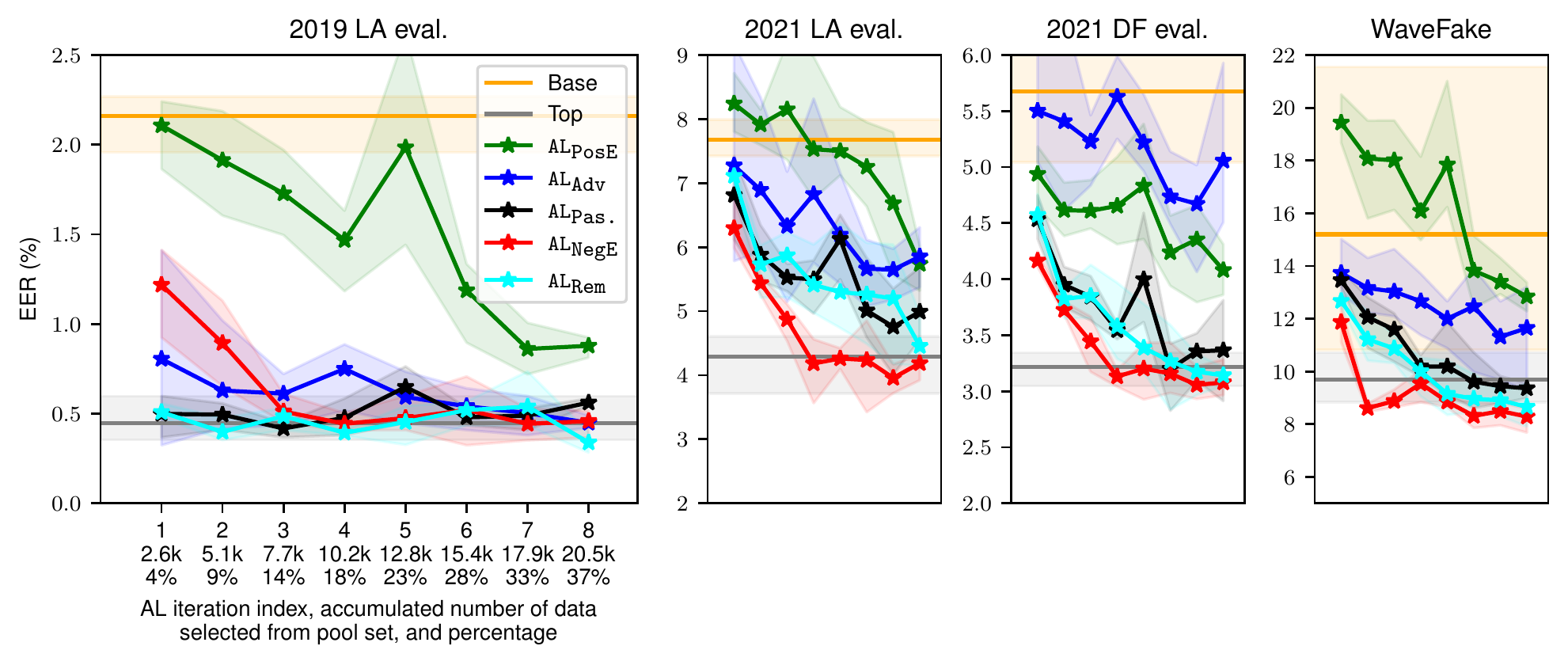}
                \caption{Results with pool set $\setD$. Note that, after eight AL iterations, 37\% of data in pool set $\setD$ were added to training set.}
        \label{fig:results_set2}
     \end{subfigure}
     \vspace{-2mm}
     \caption{Evolution of EERs (\%) on multiple test sets. Three rows of numbers below horizontal axis are AL iteration index, number of data selected from pool set, and its percentage, respectively. Each solid line represents mean EER of one CM over three runs. Upper and lower boundaries of shaded area are minimum and maximum EERs over three runs.}
     \vspace{-2mm}
     \label{fig:results}
\end{figure*}

\begin{table}[!t]
\caption{EERs (\%) on test sets. EERs of AL-based CMs were computed after last AL iteration.
EER of each CM is averaged over three runs. Lowest EER in each column and those not statistically significantly different are shown in bold font.  \BLUE{Statistical analysis on EER was conducted using $z$-statistics \cite{bengio2004statistical} and Bonferroni-Holm correction}. }
\vspace{-5mm}
\begin{center}
\footnotesize
{
\begin{tabular}{cccccc}
\toprule
 \shortstack{Pool \\ set }&              &   \shortstack{2019 \\ LA eval.}    &   \shortstack{2021 \\ LA eval.} & \shortstack{2021 \\ DF eval.} &  \shortstack{Wave \\Fake}  \\ 
\midrule
N/A  & \baseline  &     2.16     &     7.67     &     5.68     &     15.20   \\ 
  \midrule
\multirow{6}{*}{$\setA$}  
  &     \mIIIshort &     \textbf{0.23}     &     4.05     &     3.30     &     14.76   \\   
  &   \mIIshort   &     \textbf{0.22}     &     \textbf{3.96}     &     3.71     &     16.31   \\ 
  &   \mIVshort   &     \textbf{0.24}     &     4.46     &     3.78     &     16.63   \\ 
     &   \mIshort   &     \textbf{0.23}     &     \textbf{3.97}     &     3.49     &     17.94   \\ 
&   \mVshort    &     \textbf{0.20}     &     \textbf{3.73}     &     \textbf{3.28}     &     16.21   \\ 
  &   \topline    &     \textbf{0.25}     &     4.34     &     3.91     &     16.39   \\ 
  \midrule
\multirow{6}{*}{$\setD$}   
  &   \mIIIshort  &     0.88     &     5.74     &     4.08     &     12.84   \\ 
  &   \mIIshort   &     0.45     &     5.85     &     5.06     &     11.65   \\ 
  &    \mIVshort  &     0.56     &     4.99     &     3.37     &     \phantom{0}9.36    \\ 
  &   \mIshort   &     0.46     &     4.19     &     \textbf{3.08}     &     \phantom{0}\textbf{8.29}    \\ 
  &  \mVshort    &     \textbf{0.34}     &     4.45     &     \textbf{3.14}     &     \phantom{0}\textbf{8.66}    \\ 
  &   \topline    &     0.45     &     4.29     &     \textbf{3.22}     &     \phantom{0}9.69    \\ \bottomrule 
\end{tabular}
}
\vspace{-8mm}
\end{center}
\label{tab:table-eer}
\end{table}%

\begin{figure*}[t]
       \centering
       \begin{subfigure}[t]{1.1\columnwidth}
        \includegraphics[height=6cm]{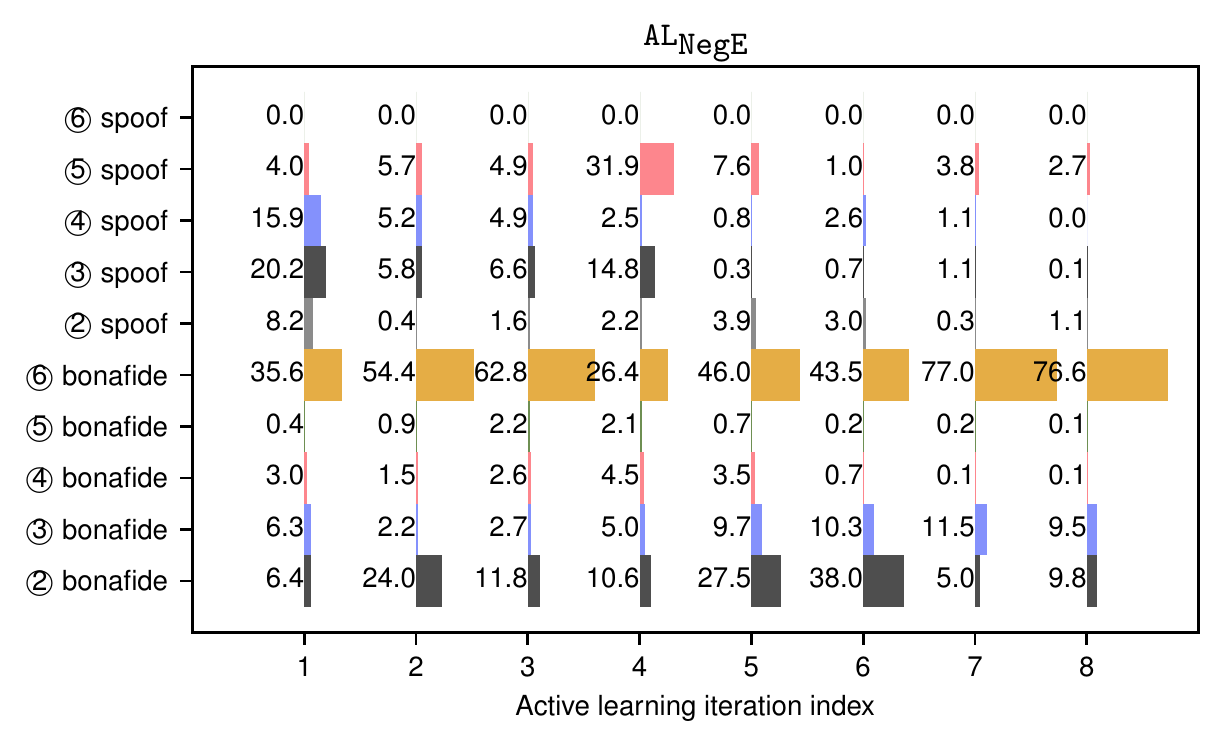}
        \label{fig:dist_0}
     \end{subfigure}
      \begin{subfigure}[t]{0.9\columnwidth}
        \includegraphics[height=6cm]{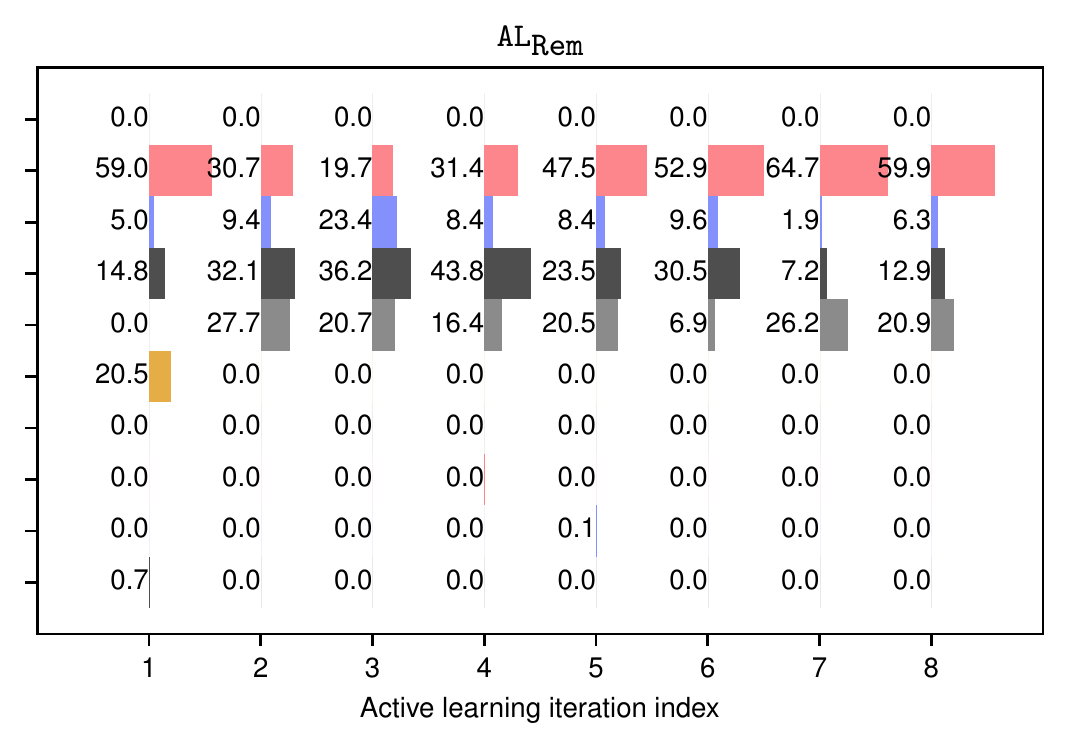}
        \label{fig:dist_4}
     \end{subfigure}
     \vspace{-5mm}
     \caption{Distribution (\%) of data selected by \mIshort\ or removed by \mVshort\ from pool set $\setD$ in each AL iteration. Sum of numbers in each iteration is equal to 100\%.}
     \vspace{-3mm}
     \label{fig:dist}
\end{figure*}

\subsection{\BLUE{Experiments and Results}}
Following the AL literature, we computed the EERs for the AL-based CMs after each AL iteration. Figure~\ref{fig:results} (a) and (b) plot the EER curves over the eight iterations when using pool set $\setA$ and $\setD$, respectively. 
For easy comparison, we also list the EERs after the last AL iteration in Table~\ref{tab:table-eer}.
Note that the AL-based CMs in Table~\ref{tab:table-eer} were trained using the same amount of data, but each CM selected different data from the pool. 
Our findings are detailed below.

\vspace{2mm}
\noindent \textbf{Does pool set affect AL performance?} The experiment results suggest ``yes''. 
For example, the results for the WaveFake test set showed that all the AL-based CMs had increased EERs as more data were selected from pool set $\setA$. In contrast, these CMs obtained improved EERs when using pool set $\setD$. 
A possible reason is that a larger and more diverse pool may contain better and more useful data.

However, a larger pool may also contain more data that are useless, and the AL-based CM has to use a proper scoring method in order to select the useful ones. 
This may explain why the differences across AL-based CMs for pool set $\setD$ were larger than those for pool set $\setA$.  As an extreme example, for the reference CM  \mIIIshort\ that only selected useless data, its EER curves when using pool set $\setD$ (Figure~\ref{fig:results} (b)) were in general higher than those for pool set $\setA$ (Figure~\ref{fig:results} (a)). 

Note that the EER curves of \mIIIshort\ for pool set $\setA$ converged to points similar to those of the other AL-based systems because the whole pool set was consumed for CM training in the last AL iteration no matter how \mIIIshort\ selected the data in previous iterations. 

\vspace{2mm}
\noindent \textbf{Which AL-based CM is more effective?} 
For pool set $\setD$, Figure~\ref{fig:results} shows that \mIshort, \mIVshort, and \mVshort\ achieved better results than the other AL-based CMs. More specifically, Table~\ref{tab:table-eer} shows that \mIshort\ and \mVshort\ achieved the top performance (i.e., no statistically significant difference from the lowest EER) on all the test sets except 2021 LA eval. Both outperformed the baseline with an at least 40\% relative reduction in EER. Furthermore, for the challenging 2021 DF and WaveFake test sets, both methods performed no worse than \topline\ while using only 37\% of the pool set for CM training.  
It should also be pointed out that  \topline\ is not a theoretical upper bound, and AL-based CMs may perform better if the pool set itself is biased \cite{zhu2017generative}.

Although {\mIshort} and {\mVshort} performed similarly at the last iteration, Figure~\ref{fig:results} shows that {\mVshort}'s EER curves dropped at a slower speed than {\mIshort} on the 2021 LA, 2021 DF, and WaveFake test sets. {\mIshort} may be preferred over the {\mVshort} if one has to choose only one method.

On pool set $\setA$, the difference across CMs was smaller. However, \mVshort\ achieved the lowest EER on all test sets except WaveFake. It performed no worse than the \topline\ while being more data efficient. 
While we do not intend to compare it with the CMs in other pieces of literature, the 0.20\% EER achieved by \mVshort\ on the 2019 LA test set was the lowest number as far as we know. 

The simple \mIVshort\ performed reasonably well, which is consistent with the findings for image and language processing tasks \cite{ducoffe2018adversarial, chandra2021initial}. In contrast, \mIIshort\ performed worse than \mIVshort\ despite being more complicated. It is not recommended for AL-based CM training.

\vspace{2mm}
\noindent \textbf{What kind of data was selected?}
Figure~\ref{fig:dist} shows the candidate data selected by \mIshort\ from pool set $\setD$ after each AL iteration. We observed that the majority was bona fide data, especially data from VoxCeleb (\textcircled{6}). 
Similarly, \mVshort\ actively removed many spoofed utterances from the pool set, especially those from BC 2019 (\textcircled{5}). After the 1st iteration, none of the bona fide data were treated as useless data or removed from the pool set. One hypothesis is that, since the ratio of bona fide versus spoofed data in the seed set was around 1:9, the AL framework tried to increase the amount of bona fide data to balance the expanded training set.

\section{Conclusion}
\label{sec:con}
This study explored AL to expand the training set for speech spoofing CMs. 
As an initial investigation, this study focused on the pool-based AL framework and compared a few existing methods to estimate the usefulness of data in a pool. A variant AL framework was also proposed that actively removes useless data from a pool. 
Among the investigated AL-based CMs, one using a negative-energy-based scoring method and one using the proposed active data-removing method achieved better performance than the others. 
Given a large and diverse pool set, both methods outperformed the strong baseline by a large margin on multiple test sets. CMs using the data set extended by the methods resulted in lower EERs not only on the ASVspoof 2019 test set but also on other test sets such as ASVSpoof 2021 and WakeFake. 
Compared with the naive method that blindly uses all data available, the two AL-based CMs achieved better or similar EERs while using less training data. Further analysis demonstrated that they prioritized bona fide data when selecting data from pool set.

Using external data to train a spoofing CM is not common because it hinders the comparison of model performance across studies. However, it is also crucial to go beyond the protocol of standard databases and explore how to boost the model performance. AL is one potential direction. Meanwhile, implementing AL for a new task is not effortless. Even using the passive learning framework requires careful selection of multiple hyper-parameters such as the candidate batch size. 
We sincerely hope that the findings and configurations presented in this study will stir new ideas on better generalizable CMs. The code, recipe, and pre-trained models will be released on https://github.com/nii-yamagishilab/project-NN-Pytorch-scripts.

\bibliographystyle{IEEEbib}
\bibliography{library}

\end{document}